\documentclass[prd, twocolumn, nofootinbib, floatfix]{revtex4-1}
\usepackage{amsmath}
\usepackage{graphicx}
\usepackage{dcolumn}
\usepackage{bm}
\usepackage{epsfig}
\usepackage{amssymb,latexsym,mathrsfs}
\usepackage{graphicx}
\usepackage{color}
\usepackage{hyperref}

\hypersetup{
    colorlinks=true,
    linkcolor=red,
    citecolor=blue,
} 

\newcommand{\be}{\begin{equation}}
\newcommand{\ee}{\end{equation}}
\newcommand{\bs}{\begin{split}} 
\newcommand{\bea}{\begin{eqnarray}}
\newcommand{\eea}{\end{eqnarray}}

\newcommand{\om}{\Omega_m}

\newcommand{\sig}{\sigma}

\newcommand{\figm}{f_{\rm IGM}} 
\newcommand{\ahe}{A_{He}}

\begin{document}

\title{Detecting Helium Reionization with Fast Radio Bursts} 

\author{Eric V.\ Linder${}^{1,2}$} 
\affiliation{
${}^1$Berkeley Center for Cosmological Physics \& Berkeley Lab, 
University of California, Berkeley, CA 94720, USA\\ 
${}^2$Energetic Cosmos Laboratory, Nazarbayev University, 
Nur-Sultan, Kazakhstan 010000}

%%%%%%%%%%%%%%%%%%%%%%%%%%%%%%%%%%%%%%%%%%%%%%%%%%%%%%%%%%%%%%%%%%%%%%%%
\begin{abstract} 
Fast radio bursts (FRB) probe the electron density of the universe along the 
path of propagation, making high redshift FRB sensitive to the helium 
reionization epoch. We analyze the signal to noise with which a detection 
of the amplitude of reionization can be made, and its redshift, for various 
cases of future FRB survey samples,  
assessing survey 
characteristics including total number, redshift distribution, peak 
redshift, redshift depth, and number above the reionization redshift, 
as well as dependence on reionization redshift. 
We take into account scatter in the 
dispersion measure due to an inhomogeneous intergalactic medium (IGM) and 
uncertainty in the FRB host and environment dispersion measure,  as   well as  
cosmology. For a future 
survey with 500 FRB extending out to $z=5$, and a sudden reionization, the 
detection of helium reionization can approach the 
$5\sigma$ level and 
the reionization redshift be determined to $\sigma(z_r)\approx0.24$ in 
an optimistic scenario, or $2\sigma$ and $\sigma(z_r)\approx0.34$ taking 
into account further uncertainties on IGM fraction evolution and redshift 
uncertainties. 
\end{abstract}

\date{\today} 

\maketitle

%%%%%%%%%%%%%%%%%%%%%%%%%%%%%%%
\section{Introduction}

Reionization is one of the most dramatic processes in the young universe, 
turning the cosmic medium from its long, substantially neutral condition into 
an ionized state. For hydrogen atoms this occurs in the first billion years 
of the age of the universe, at redshifts $z\gtrsim6$, and is a subject of  
extremely active research. A later epoch of reionization, more accessible 
to direct measurements by next generation surveys, is that of helium. While 
neutral helium loses its outer electron early, He II is believed to have its 
more tightly bound inner electron ionized when the universe is some two 
billion years old, at $z\approx3$. 

We refer to this later process simply 
as helium reionization, and relatively little 
is known about exactly when it occurs. Measurements of the event could shed 
light on energetic processes during this important epoch of the universe, 
when quasar black hole activity and star formation is nearing its peak. 
The addition of further free electrons to the intergalactic medium increases 
the optical depth to electromagnetic waves, and can be probed for example 
through the dispersion measure of fast radio bursts (FRB). The higher the 
electron density, the higher the plasma frequency for radio wave propagation,  
and the greater the frequency dispersion -- quantified by the dispersion 
measure (DM) as an integral along the propagation path. 
For recent reviews of FRB, see \cite{1906.05878,1904.07947}. 
A very partial sample of some 
recent articles using FRB to learn about astrophysics and cosmology 
includes \cite{1909.02821,kumlin,1903.06535,1811.00899,1811.00197}. 

Here we investigate guidelines and early steps for assessing 
the ability of future FRB surveys to high redshift to 
detect helium reionization and estimate its characteristics. 
Section~\ref{sec:method} describes the method and definition of a signal to 
noise for detection of the occurrence of helium reionization. We study the 
dependence of the results on survey characteristics such as number of 
sources, survey depth, redshift distribution, redshift peak, and number 
above the reionization redshift in Sec.~\ref{sec:results}, as well as 
the role of IGM inhomogeneity and host galaxy DM uncertainty. 
The impact of further astrophysical or measurement 
uncertainties is considered in Sec.~\ref{sec:uncert}. We 
conclude in Sec.~\ref{sec:concl}.

%%%%%%%%%%%%%%%%%%%%%%%%%%%% 
\section{Dispersion Measure and Helium Reionization} \label{sec:method} 

We begin with a brief review of the key cosmological and astrophysical 
quantities entering the FRB measurements. 

Fast radio bursts exhibit a characteristic sweep of the signal frequency 
with time, due to its propagation through an ionized medium. This can be 
characterized by its dispersion measure and interpreted as the integral 
along the line of sight of the electron density from emitter to observer. 
Thus, 
\be 
{\rm DM}=\int_{t_e}^{t_o} dt\,n_e(t)\,[1+z(t)]^{-1}\ , 
\ee 
where the $1+z$ factor arises from frequency and time transformations 
between emitter and observer frames. The units are conventionally given  
as parsecs/cm$^3$, though we will suppress explicitly writing them, as 
well as setting the speed of light to one. 

The propagation time can be written in terms of the cosmological expansion 
rate, or Hubble parameter $H(z)$, as $dt=dz/[(1+z)H(z)]$, where $z$ is 
the redshift. Thus the cosmological model enters through $H(z)$, and if 
the electron density is well known then one could attempt to use FRB to 
constrain cosmology (e.g.\ 
\cite{1903.06535,1901.02418,1812.11936,1711.11277}, including a particular 
analysis of the effects of systematic uncertainties on cosmological 
parameter determination in \cite{kumlin}). 
Here we take the converse approach and seek to 
learn about the cosmic ionization history through $n_e(z)$ 
(see also \cite{1808.00908,1902.06981}). 

If the universe were perfectly homogeneous, and completely ionized for all 
redshifts of interest, then $n_e(z)=n_{e,0}(1+z)^3$ along every line of 
sight. However, the intergalactic medium is in fact inhomogeneous, giving 
a scatter to $n_e(z)$ at any specified redshift, and we are looking for 
signs of the helium reionization. Therefore we write 
$\langle n_e(z)\rangle=n_{e,0}(1+z)^3 f_e(z)$ and consider the ionization 
history $f_e(z)$, while including variance in $n_e(z)$ at each redshift 
in the error budget. 

We adopt the standard model for the ionization fraction, taking the 
universe to be made of a mass fraction $Y$ of helium and $1-Y$ of hydrogen. 
Since we will be considering surveys out to redshifts $z<6$, then we take 
all of the hydrogen to be ionized and none of the helium to be neutral 
(i.e.\ it is either singly ionized He II or fully ionized He III). In 
summary, 
\be 
f_e(z)=(1-Y)f_H+\frac{Y}{4}(f_{\rm He II}+2f_{\rm He III})\ , \label{eq:fe}
\ee 
with the 2 accounting for He III having released 2 electrons to the cosmic 
medium. Thus, $f_H=1$, and before helium reionization $f_{\rm He II}=1$ 
and $f_{\rm He III}=0$, while afterward $f_{\rm He II}=0$ and 
$f_{\rm He III}=1$. Thus, $f_e(z>z_r)=1-3Y/4=0.818$ and 
$f_e(z<z_r)=1-Y/2=0.879$, where we take the Planck value of $Y=0.243$ 
\cite{planckhe}  (also see \cite{aver}). 
The ionization fraction increases by $\approx7.5\%$ due 
to helium reionization. We assume the reionization occurs suddenly at  
redshift $z_r$ (though we later explore variations of this). 

The dispersion measure due to cosmology (i.e.\ the propagation through the 
intergalactic medium) is therefore 
\be 
{\rm DM}=H_0^{-1}n_{e,0}\int_{z_e}^{z_o} \frac{dz\,(1+z)}{H(z)/H_0}\,f_e(z)\ . 
\label{eq:dmcos}  
\ee 
The host galaxy of the FRB and its local environment also contribute to 
the detected dispersion measure. This is suppressed by a factor $1+z$, but 
we take it into account in our error budget. The Milky Way galaxy also 
gives rise to dispersion, but good maps exist to subtract this out 
\cite{ymw,cordes}. Therefore we take the ``measured'' value to be compared  
to theory to be simply given by Eq.~(\ref{eq:dmcos}). 

The first question concerning helium reionization is whether we can detect 
it in a sample of high redshift FRB. Thus we want to explore a simple 
signal to noise criterion for detection. Beyond that we want to know when 
it occurs, i.e.\ estimate $z_r$. To accomplish this we adopt the form 
\be 
{\rm DM}(z)={\rm DM}(z)_{\rm high}+\ahe 
\left[{\rm DM}(z)_{z_r}-{\rm DM}(z)_{\rm high}\right]\ . \label{eq:dmform} 
\ee 
Here the subscript ``high'' means reionization occurs beyond the limits 
of the sample, while the subscript ``$z_r$'' means it occurs at $z=z_r$. 
If $\ahe=0$ then reionization is not detected by the sample, while if 
$\ahe=1$ it is consistent with reionization occurring at $z_r$. Thus, if 
we determine $\ahe$ to $\sigma(\ahe)$ then the detection signal to noise 
is $\ahe/\sigma(\ahe)$, or simply $1/\sigma(\ahe)$ since $\ahe=1$ if the 
model is correct. 

\begin{figure}[tbp!]
\includegraphics[width=\columnwidth]{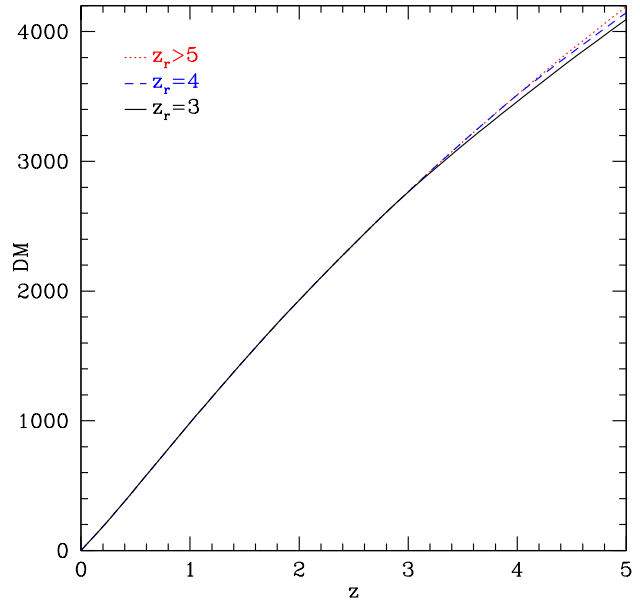}
\caption{
The cosmological DM has a clear change in slope at the reionization 
redshift. While the model with reionization occurring above the sample 
range (dotted red, with $z_r>5$) shows no special bend, the baseline 
model with $z_r=3$ (solid black) does. We also plot a case with higher 
than expected reionization redshift, $z_r=4$ (dashed blue). 
} 
\label{fig:dmtot} 
\end{figure}

Figure~\ref{fig:dmtot} illustrates the model and its effect on the 
cosmological DM. The cosmological DM has a significant change in slope with 
redshift due to the extra influx of electron density from  helium 
reionization. The key questions are, first, can we detect this signature, 
i.e.\ find $\ahe\ne0$ with significant signal to noise, and can we measure 
the redshift $z_r$ at which reionization occurs. While the model curves 
appear close to each other, the signal to noise is enhanced by having many 
FRB over a wide redshift range on either side of the reionization redshift.

%%%%%%%%%%%%%%%% 

We use the Fisher information matrix formalism to estimate the parameters 
$\ahe$ and $z_r$, and include the cosmological parameter $\om$ (we assume 
a flat $\Lambda$CDM cosmology, where $H(z)/H_0$ is fully characterized by 
$\om$, the matter density as a fraction of the critical density). The 
fiducial values are taken to be $\om=0.3$, $\ahe=1$, and $z_r=3$, though 
we will study the effect of different fiducial $z_r$ in the next section. 
Note that for the Fisher formalism the logarithmic derivatives of the 
measured quantity with respect to the fit parameters are the key elements, 
when the measurement errors dominate as we take here, so the constant 
prefactor in Eq.~(\ref{eq:dmcos}) is not important. 

For the mock data, we use 500 FRB from a future survey, with redshifts, 
reaching out to 
$z_{\rm max}=5$. Our baseline FRB number density distribution with redshift 
increases at low redshift as the cosmic volume increases, then peaks around 
the expected helium reionization redshift, and declines at higher redshift 
due to diminished signal to noise. It is intended as a toy model, not a 
real survey  distribution which would depend on detailed modeling of 
instrumental characteristics. To test sensitivity of the results to the 
survey we also compare to a distribution uniform with redshift; we do not 
find significant sensitivity to the exact form of the distribution. In 
Sec.~\ref{sec:results} we also study the effect of variation of $z_{\rm max}$ 
and the peak redshift. In Sec.~\ref{sec:uncert} we discuss redshift 
uncertainties. 

The error model for the DM ``cleaned'' measurements (i.e.\ with the 
contributions from within the Milky Way and host galaxy removed) 
includes scatter due 
to the inhomogeneous intergalactic medium (IGM), using the form \cite{kumlin} 
\be 
\frac{\sigma({\rm DM})_{\rm IGM}}{{\rm DM}_{\rm IGM}}=\frac{20\%}{\sqrt{z}} 
\ , \label{eq:prec} 
\ee 
inspired by  simulations \cite{1309.4451,1812.11936,1712.01280,1901.11051}. 
The host  and local environment contributions to the detected DM may be 
random among the FRB sample; we do not actually need the value of the mean 
of those noncosmological fluctuations 
for our analysis (i.e.\ the value that will be subtracted off to define the 
cosmological contribution) only the residual scatter that will presumably 
be random and hence reduced by $\sqrt{N}$ using $N$ FRB. This is not well 
known; we assume a residual 
uncertainty $\sigma({\rm DM})_{\rm host}\approx 30$ 
(redshift independent, i.e.\ $30(1+z)$ in the host frame, reflecting greater 
uncertainty at higher redshift), but study the effects of varying this. 

\begin{figure}[tbp!]
\includegraphics[width=\columnwidth]{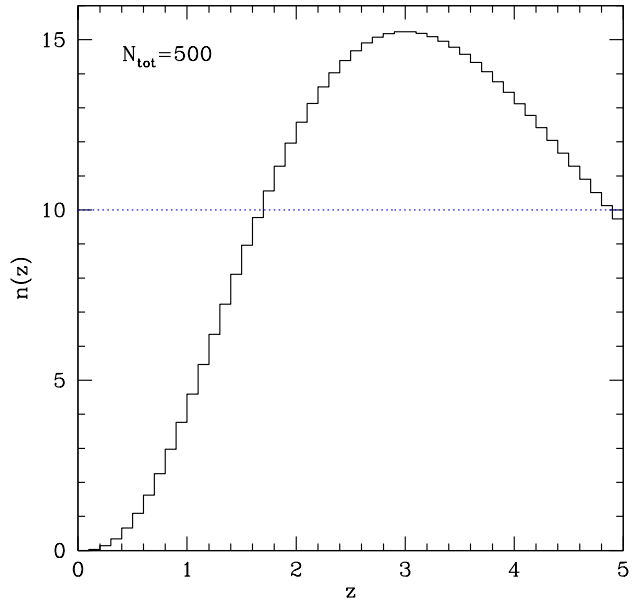}
\caption{
The mock FRB sample distribution is shown in redshift bins of width 0.1  
for the baseline  (solid black) and uniform (dotted blue) distributions. 
Both contain 500 FRB. 
} 
\label{fig:nz} 
\end{figure}

Figure~\ref{fig:nz} shows the baseline and uniform distributions of FRB 
in our mock surveys. The baseline form is  
\be 
n_{\rm FRB}(z)\sim z^3\,e^{-z/z_\star}\ , \label{eq:nz}
\ee 
and the uniform one is $n(z)=\,$const, both normalized to give 500 FRB 
from $z=[0,z_{\rm max}]$. Fiducial values are $z_{\rm max}=5$, $z_\star=1$, 
though we later vary these.

%%%%%%%%%%%%%%%% 

\begin{figure}[tbp!]
\includegraphics[width=\columnwidth]{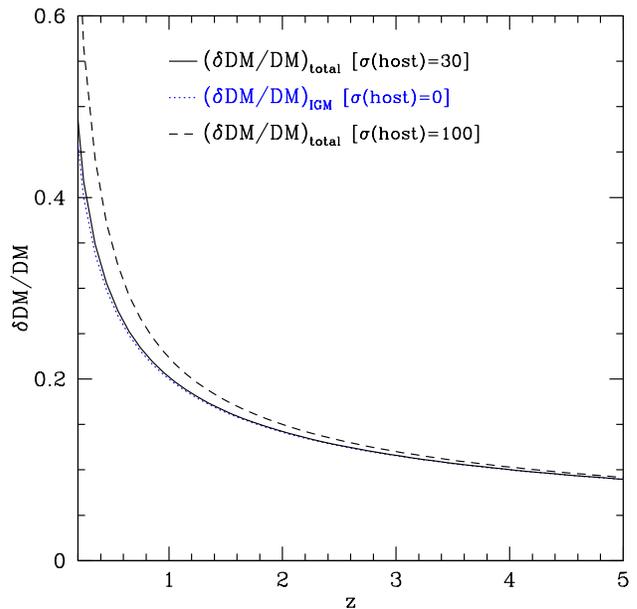}
\caption{
The fractional uncertainty contributing to a measured DM, due to IGM scatter 
and uncertainty in the host and local environment contribution, is plotted 
vs redshift. The quadrature sum of the two contributions is plotted as the 
solid black curve, using a baseline host uncertainty $\delta{\rm DM}=30$, 
while using an increased host uncertainty of 100 is shown by the dashed black 
curve. The dotted blue curve gives the IGM scatter contribution alone. In 
the high redshift region with the most information on helium reionization 
the IGM component dominates. 
} 
\label{fig:uncert} 
\end{figure}

Figure~\ref{fig:uncert} shows the model for the 
fractional uncertainty on an individual cosmological DM 
measurement, accounting for scatter in IGM properties and host and local 
environment variations. Since DM is small at low redshift, the fractional 
uncertainty is large there. At higher redshift, lines of sight average over 
IGM inhomogeneity, and DM increases, so the fractional uncertainty decreases. 
The host and local environment residual contribution is much smaller than 
the IGM uncertainty -- for example at $z=1$ the IGM DM scatter is 
$\approx 210$, so adding the host uncertainty of 30 (not DM$_{\rm host}$) 
in quadrature gives 
negligible increase. We also show the effect of increasing the host 
uncertainty to 100; it is still not a large effect, especially near the 
expected reionization redshift $z_r\gtrsim3$.

%%%%%%%%%%%%%%%% 

%%%%%%%%%%%%%%%%%%%%%%%%%%%%  
\section{Results and Survey Dependence} \label{sec:results} 

Having all the elements in place, we can now use the Fisher information 
analysis to evaluate the parameter estimation. For our baseline case we 
find the marginalized parameter uncertainties are $\sig(\ahe)=0.22$ and 
$\sig(z_r)=0.24$. That is, the signal to noise for detection, 
$S/N=\ahe/\sig(\ahe)=1/\sig(\ahe)=4.5$. Furthermore, we can constrain the 
sudden reionization model to have $z_r=3\pm0.24$. Since all the input 
errors are treated statistically, all uncertainties will scale as 
$1/\sqrt{N}$. (Recall we use $N=500$.) 

\begin{figure}[tbp!]
\includegraphics[width=\columnwidth]{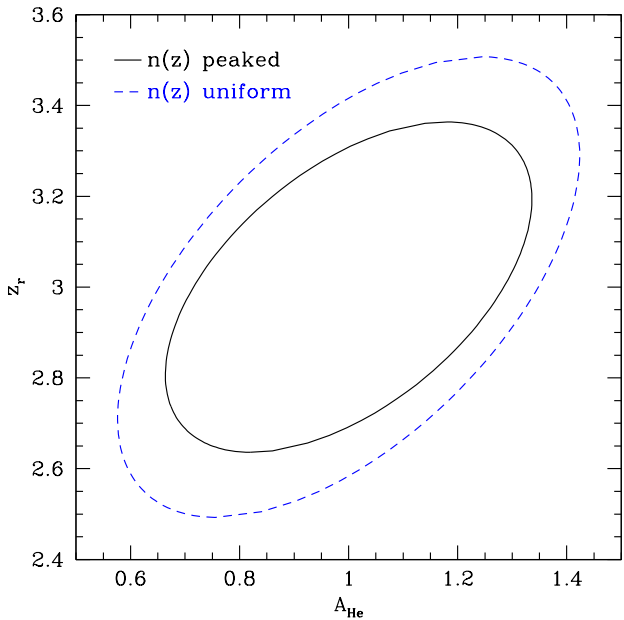}
\caption{
68.3\% ($1\sig$) joint confidence contours on the reionization parameters 
$\ahe$ and $z_r$ are plotted for the baseline FRB distribution 
(``peaked'': solid black) and uniform distribution (dashed blue). Such 
FRB samples would clearly enable distinction from $\ahe=0$, i.e.\ detection 
of helium reionization within the sample redshift range, and reasonable 
localization of $z_r$. 
} 
\label{fig:ellaz} 
\end{figure}

There is not a strong covariance between $\ahe$ and $z_r$ determinations, 
and they are not especially sensitive to the difference between the 
baseline and uniform FRB distributions. (At the end of this section we also 
investigate further the role of $z_{\rm max}$ and the distribution peak 
$z_{\rm peak}=3z_\star$.) For example, the parameter uncertainties 
increase to 0.28 and 
0.33, respectively. Figure~\ref{fig:ellaz} shows the 68.3\% ($1\sigma$) 
joint confidence contours.

%%%%%%%%%%%%%%%% 

\begin{figure}[tbp!]
\includegraphics[width=\columnwidth]{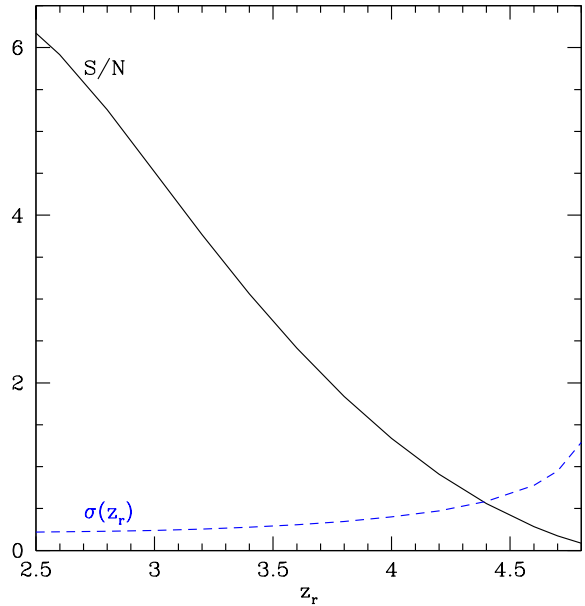}
\caption{
The signal to noise $S/N=1/\sig(\ahe)$ for detection of reionization 
and the uncertainty on the reionization redshift $\sig(z_r)$ are plotted 
vs the value of $z_r$. As one approaches the sample edge at $z_{\rm max}=5$, 
the S/N vanishes and the redshift uncertainty blows up. 
} 
\label{fig:varyr} 
\end{figure}

We can explore the sensitivity as we vary the reionization redshift. 
Figure~\ref{fig:varyr} shows that the S/N increases appreciably for lower 
$z_r$, but fades below $S/N=1$ for $z_r\gtrsim4$. The uncertainty 
$\sig(z_r)$ has a slower dependence, though it becomes steeper as one 
approaches the edge of the sample. Over the main part of the expected 
range, say $z_r\approx2.5$--3.5, the S/N varies between 6.2 and 2.7, and 
$\sig(z_r)$ between 0.22 and 0.29.

%%%%%%%%%%%%%%%% 

If we assume our model is correct, i.e.\ sudden reionization at $z_r$ 
does occur within the sample range, and fix $\ahe=1$, then the uncertainty  
$\sig(z_r)$ only decreases from 0.24 to 0.20 for the fiducial case $z_r=3$. 
This is due to the lack of strong covariance between $\ahe$ and $z_r$. 

One could also relax the assumption of sudden reionization and attempt 
to see a signal of a more gradual transition. This will correspond to 
finding (e.g.\ within a Monte Carlo parameter estimation) a fit value of 
$\ahe\ne1$. Using Eqs.~(\ref{eq:fe})--(\ref{eq:dmform}) one finds that 
one can write 
\be 
\ahe(z)=1-\frac{\int_{z_r}^z \frac{dz'\,(1+z')}{H(z')/H_0}\,f_{\rm HeIII}(z)}{\int_{z_r}^z \frac{dz'\,(1+z')}{H(z')/H_0}}\ . 
\ee 
That is, $\ahe$ is (one minus) a weighted ionization fraction. For sudden  
reionization this reduces to what we use throughout this article. Thus a 
fit value $\ahe$ statistically distinct from one or zero would point to 
a gradual reionization process, where $f_{\rm HeIII}$ does not suddenly  
transition from zero to one. That serves as an alert that, while reionization 
is detected, one should carry out an analysis with a redshift dependent 
function rather than a single constant parameter $\ahe$ (for example a 
principal components analysis as in \cite{hu1,hu2}). 

We can check the robustness of the results for other variations in the 
inputs. If we change the host uncertainty contribution from 30 to 10 or 100, 
then $\sig(\ahe)$ goes from 0.22 to 0.22 or 0.23, respectively, and 
$\sig(z_r)$ from 0.24 to 0.24 or 0.25. 

\begin{figure}[tbp!]
\includegraphics[width=\columnwidth]{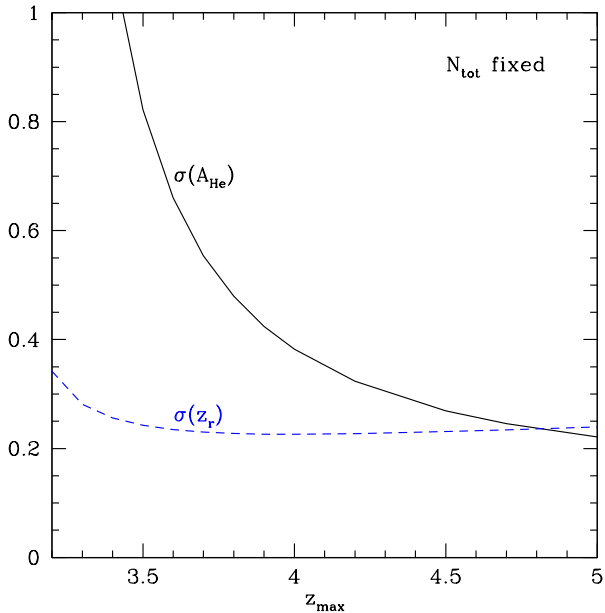}
\caption{
Changing the depth of the FRB sample, $z_{\rm  max}$, has a significant 
effect on the inverse S/N, $\sig(\ahe)$ (solid black curve), but less on 
the reionization redshift uncertainty (dashed blue). Here we hold the 
total number of FRB fixed as we change $z_{\rm max}$, keeping the form 
of the redshift distribution the same. 
} 
\label{fig:zmax} 
\end{figure}

Increasing the sample depth $z_{\rm max}$ will certainly decrease the 
parameter uncertainties, but it is fairer to do so while holding the 
total number of FRB fixed. As seen in Fig.~\ref{fig:zmax}, in 
this case the S/N gets worse rapidly with 
decreased depth, due to the reduced lever arm above the reionization 
redshift. However the localization of the bend in DM, i.e.\ $\sig(z_r)$ 
stays fairly constant until the sample depth approaches $z_r$ (where 
one loses leverage above $z_r$).

%%%%%%%%%%%%%%%% 

We can also change how the FRB distribution out to the fiducial 
$z_{\rm max}=5$ is shaped. The peak of the distribution in Eq.~(\ref{eq:nz}) 
occurs at $z_{\rm peak}=3z_\star$. We can either keep the total number 
of FRB fixed, or the number of FRB with $z>z_r$ fixed. Our fiducial case 
has 500 total FRBs, of which 260, or just about half are at $z>z_r$. Fixing 
$N(z>z_r)$ is interesting since we might expect the leverage on the 
reionization parameters to come from the different behavior at $z>z_r$ before 
reionization, i.e.\ DM($z<z_r$) does not depend on $\ahe$ or $z_r$. 

\begin{figure}[tbp!]
\includegraphics[width=\columnwidth]{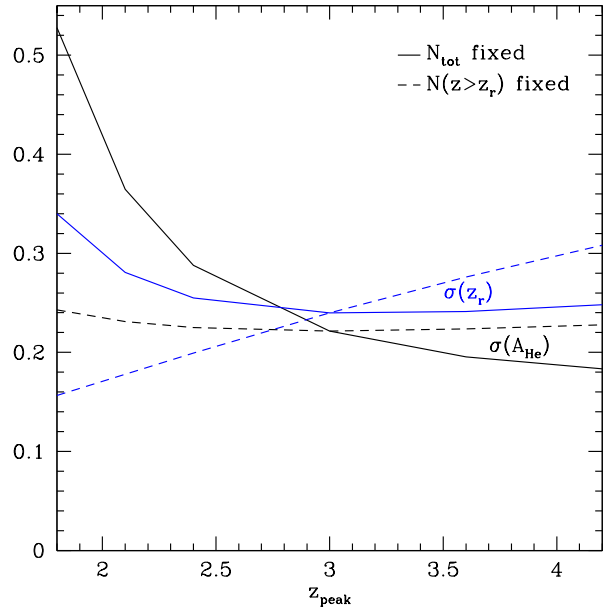}
\caption{
Changing the shape of the FRB distribution, here quantified in terms of 
the peak redshift, does not have a large effect 
on the parameter estimation ($\ahe$ in black, $z_r$ in light blue) 
as long as the range on either side 
of the reionization redshift is reasonably sampled. Solid curves show 
the case where the total number of FRB in the full distribution is held 
fixed, while dashed curves preserve the number of FRB above the reionization 
redshift. 
} 
\label{fig:zstar} 
\end{figure}

Figure~\ref{fig:zstar} shows that this is indeed so. If we fix $N_{\rm tot}$, 
then moving the peak of the distribution to lower redshifts worsens the 
uncertainty on $\ahe$, i.e.\ it becomes harder to be sure that reionization 
occurred at all (the S/N drops below 2 for $z_{\rm peak}=1.8$; note there  
are then only 120 FRB above $z_r$). However, 
fixing the number above $z_r$ instead greatly ameliorates this effect, 
nearly leveling out $\sig(\ahe)$ despite the change in  $z_{\rm peak}$. 
For $\sig(z_r)$, however, while peaking the distribution at low redshift 
worsens its estimation, this is completely overturned when fixing $N(z>z_r)$ 
because there are also many more total FRB; for $z_{\rm peak}=1.8$ there 
are then 1087 total FRB in the full distribution.

%%%%%%%%%%%%%%%% 

%%%%%%%%%%%%%%%%%%%%%%% 
\section{Further Uncertainties} \label{sec:uncert} 

In using Eq.~(\ref{eq:dmcos}) to account for the effect of helium 
reionization on the electron number density, we assumed that all free 
electrons 
contributed to the intergalactic medium. However, this is not completely 
true and one often includes a factor $\figm$ in the equation representing 
the fraction present in the IGM. To the extent that this is constant with 
redshift, it can be absorbed in the other constant factors and does not 
affect the results so far. However, a modest ($\sim20\%$) evolution is 
expected over the range $z=0$--5 (see, e.g., 
\cite{1309.4451,1901.11051,1909.02821}), though 
slower around the range of redshifts 
of interest for helium reionization \cite{shull12,1812.11936}. We here write 
\be 
\figm(z)=\figm(z=3)\,[1+s(z-3)]\ , 
\ee 
and add $s$ as a fit parameter. 

\begin{figure}[tbp!]
\includegraphics[width=\columnwidth]{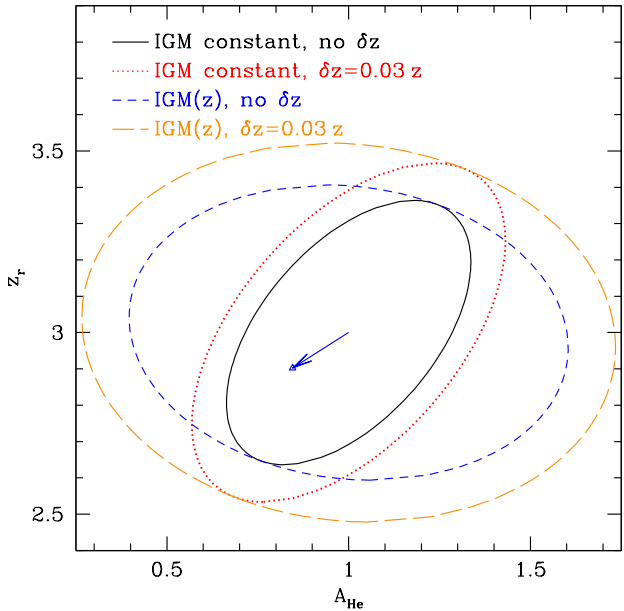}
\caption{
68.3\% ($1\sig$) joint confidence contours on the reionization parameters 
$\ahe$ and $z_r$ are plotted for the baseline FRB distribution as in 
Fig.~\ref{fig:ellaz} (solid black) and with various additional uncertainties. 
Allowing linear evolution of the IGM fraction $\figm(z)$ gives the 
dashed blue contour, while including redshift uncertainty as a contribution 
to the measurement covariance matrix yields the dotted red contour. Adding 
both gives the long dashed orange contour. Treating redshift errors as a 
systematic offset will bias the astrophysics results by changing their 
values as shown by the blue arrow giving the shift in the center of 
the dashed blue contour. 
} 
\label{fig:uncertsz} 
\end{figure}

We then incorporate the slope $s$ into the Fisher information analysis 
and assess the impact on the parameter uncertainties. 
Figure~\ref{fig:uncertsz} 
shows the results (for this case and the others in this section). 
Here we focus on the effect of including $\figm(z)$, 
corresponding to going from the solid black contour (the same as in 
Fig.~\ref{fig:ellaz}) to the dashed blue contour. 
The uncertainty on $\ahe$ increases from 0.22 to 0.40, 
changing the ``detection'' S/N from 4.5 to 2.5. Essentially the evolution 
of the IGM fraction blurs the bend due to helium reionization; of course 
for a free function $\figm(z)$ one could exactly offset the change in the 
ionization fraction.

%%%%%%%%%%%%%%%% 

Another uncertainty is the FRB redshift. Localization to a specific galaxy 
can be a challenge, especially at high redshift. One can also take a 
statistical approach by crosscorrelating with large scale structure in the 
uncertainty volume. What we are interested in 
is not individual redshifts per se, but rather the uncertainty of estimation 
over a number of FRB within a redshift bin. The central limit theorem helps 
in that even if individual redshift probability distributions may be 
non-Gaussian, the ensemble estimate should be Gaussian. We want the 
uncertainty in that estimation. This is a common issue in cosmological  
distance estimation, e.g.\ from supernovae without spectroscopic redshifts 
or gravitational wave standard sirens. 

We approach it in two ways: in the first method we include an ensemble 
redshift uncertainty in the measurement noise matrix, added in quadrature 
with the other uncertainties, while in the second method we use the Fisher 
bias formalism (see \cite{snbias} for its application to photometric 
supernovae, \cite{gwbias} for its application to standard sirens, and 
\cite{kumlin} for its application using a different uncertainty to FRB). 

For the first, statistical method we add 
\be 
\left[\delta z\frac{\partial{\rm DM}(z_i)}{\partial z}\right]^2 
\ee 
to the noise matrix, taking each bin $i$ independently. 
We adopt an uncertainty $\delta z=\alpha z$, with $\alpha=0.03$. This 
increases the uncertainty on the astrophysical parameter estimation, 
specifically diluting the $\ahe$ determination from 0.22 to 0.28, and 
$\sig(z_r)$ from 0.24 to 0.31. This is shown in Fig.~\ref{fig:uncertsz} 
for both the redshift uncertainty alone, and in combination with the 
previous IGM evolution uncertainty where the parameter uncertainties 
increase to 0.48 and 0.34 respectively. 

For the second, systematic uncertainty we instead apply $\delta z=\alpha z$, 
again with $\alpha=0.03$, as a systematic offset to each redshift bin. 
The shift in the confidence contour central values, for the case including 
IGM evolution as well, 
is denoted by the blue arrow in Fig.~\ref{fig:uncertsz}. 
We can assess the impact of the shift by taking into account the joint 
bias between $\ahe$ and $z_r$ (i.e.\ direction as well as magnitude) in 
terms of the $\Delta\chi^2$ shifted. The value for $\alpha=0.03$ (and it 
scales quadratically with $\alpha$ for small bias) is $\Delta\chi^2=0.3$. 
Since the $1\sig$ joint confidence contour lies at $\Delta\chi^2=2.3$, this 
can be thought of as a $0.14\sig$ joint shift. Thus, redshift uncertainties 
at the 0.03 level (for the redshift bin ensemble, not individual FRB) are 
not very significant, whether treated through the statistical or systematic 
approach.

%%%%%%%%%%%%%%%%%%%%%%% 
\section{Conclusions} \label{sec:concl} 

Helium reionization is an important transition in the evolution of the 
cosmic medium, and bears potential clues to quasar black hole activity, 
star formation, and baryon distribution. Fast radio bursts have, well, 
burst onto the scene as a promising tool for probing the intergalactic 
medium to these redshifts. 

Future surveys will have many thousands of 
FRB detections, and new telescopes offer the possibility of tight 
localization. While identification of counterpart galaxies, and 
redshifts, may be more challenging at high redshift, new ideas such 
as Ultra-Fast Astronomy \cite{ufa} in the optical may help. An alternative 
approach is statistical crosscorrelation with large scale structure to 
obtain redshifts. Here we considered 500 FRB with redshifts out to 
$z\approx5$, though we discussed the scaling of results with numbers and 
depth. 

We explored guidelines for where lay the greatest leverage in survey 
characteristics, assessing total number, redshift distribution, peak 
redshift, redshift depth, and number above the reionization redshift. 
The last was identified as the most important variable (though it is 
important to have a lever arm on both sides of the helium reionization 
epoch). We also examined how the value of the reionization redshift $z_r$ 
affected prospects for detection and found the effect was modest within 
the expected redshift range. 

Defining a signal to noise criterion for detecting helium reionization, 
we carried out a Fisher information analysis to estimate constraints on 
the S/N and the redshift of reionization. We included uncertainties due 
to a inhomogeneous IGM, host galaxy contributions, and cosmology (within 
$\Lambda$CDM). For 500 FRB, roughly half above and half below the reionization 
redshift, the detection S/N $\approx4.5$ and $\sigma(z_r)\approx0.24$. 
A value of the ``reionizationness'' $\ahe<1$ could be interpreted 
as a weighted reionization fraction $1-\langle f_{\rm HeIII}\rangle$ 
that indicates a more gradual transition than instantaneous reionization. 
These calculations are early steps and guidelines toward future work 
using more sophisticated simulations of IGM properties and potential 
methods for reducing uncertainty in the IGM and host contributions. 

We also investigated systematics in terms of evolution in the IGM fraction, 
finding that it could significantly weaken detection of reionization, so 
that complementary observations that could constrain this (e.g.\ through 
X-ray, Sunyaev-Zel'dovich, or neutral hydrogen measurements) could play 
an important role. Uncertainties in FRB redshifts were accounted for in 
two ways: treating them as a statistical uncertainty ``bloating'' the 
parameter estimation or as a systematic bias. In both cases, if the 
redshift uncertainty can be controlled at the $\sim3\%$ fractional level 
then parameter estimation is not seriously degraded. 

Whether obtaining such a large sample of FRB, with well estimated redshifts, 
at such high redshift is realistic or not is difficult to foretell. However 
the extraordinary rapid development of FRB detections in the last couple 
of years argues that consideration of the possibilities for such a sample 
to explore the cosmic medium should not be neglected.

%%%%%%%%%%%%% 
\acknowledgments 

I gratefully acknowledge very useful discussions with Pawan Kumar, 
supported from MES AP05135753. 
This work is supported in part by the Energetic Cosmos Laboratory and by the 
U.S.\ Department of Energy, Office of Science, Office of High Energy Physics, 
under Award DE-SC-0007867 and contract no.\ DE-AC02-05CH11231.

%%%%%%%%%%%%%%%%%%%%%%% 


\begin{thebibliography}{99}

\bibitem{1906.05878} 
J.M. Cordes, S. Chatterjee, Fast Radio Bursts: An Extragalactic Enigma, 
Ann. Rev. Astron. Astroph. 57, 417 (2019) [arXiv:1906.05878] 

\bibitem{1904.07947} 
E. Petroff, J.W.T. Hessels, D.R. Lorimer, Fast Radio Bursts, Astron. 
Astroph. Rev. 27, 4 (2019) [arXiv:1904.07947] 

\bibitem{1909.02821} 
A. Walters, Y-Z. Ma, J. Sievers, A. Weltman, Probing Diffuse Gas with Fast 
Radio Bursts, Phys. Rev. D 100, 103519 (2019) [arXiv:1909.02821] 

\bibitem{kumlin} 
P. Kumar, E.V. Linder, Use of Fast Radio Burst Dispersion Measures as 
Distance Measures, Phys. Rev. D 100, 083533 (2019) [arXiv:1903.08175] 

\bibitem{1903.06535} 
V. Ravi et al., Fast Radio Burst Tomography of the Unseen Universe, 
arXiv:1903.06535 

\bibitem{1811.00899}  
E.F. Keane, The Future of Fast Radio Burst Science, 
Nature Ast. 2, 865 (2018) [arXiv:1811.00899] 

\bibitem{1811.00197} 
J-P. Macquart, Probing the Universe's baryons with fast radio bursts, 
Nature Ast. 2, 836 (2018) [arXiv:1811.00197]

\bibitem{1901.02418} 
M.S. Madhavacheril, N. Battaglia, K.M. Smith, J.L. Sievers, 
Cosmology with kSZ: breaking the optical depth degeneracy with Fast Radio 
Bursts, Phys. Rev. D 100, 103532 (2019) [arXiv:1901.02418] 

\bibitem{1812.11936} 
M. Jaroszynski, Fast Radio Bursts and cosmological tests, MNRAS 484, 1637 
(2019) [arXiv:1812.11936] 

\bibitem{1711.11277}
A. Walters, A. Weltman, B.M. Gaensler, Y-Z. Ma, A. Witzemann, 
Future Cosmological Constraints from Fast Radio Bursts, ApJ 856, 65 (2018) 
[arXiv:1711.11277] 

\bibitem{1808.00908} 
J-P. Macquart, R.D. Ekers, FRB event rate counts II -- fluence, redshift and 
dispersion measure distributions, MNRAS 480, 4211 (2018) [arXiv:1808.00908] 

\bibitem{1902.06981} 
M. Caleb, C. Flynn, B. Stappers, Constraining the era of helium reionization 
using fast radio bursts, MNRAS 485, 2281 (2019) [arXiv:1902.06981] 

\bibitem{planckhe} 
Planck Collaboration, Planck 2018 results.\ VI.\ Cosmological parameters, 
arXiv:1807.06209 

\bibitem{aver}  
E. Aver, K.A. Olive, E.D. Skillman, The effects of He I $\lambda$10830 on 
helium abundance determinations, JCAP 1507, 011 (2015) [arXiv:1503.08146] 

\bibitem{ymw} 
J.M. Yao, R.N. Manchester, N. Wang, A New Electron Density Model for 
Estimation of Pulsar and FRB Distances, ApJ 835, 29 (2017) [arXiv:1610.09448] 

\bibitem{cordes} 
J.M. Cordes, T.J.W. Lazio, NE2001.I. A New Model for the Galactic Distribution 
of Free Electrons and its Fluctuations, arXiv:astro-ph/0207156 

\bibitem{1309.4451} 
M. McQuinn, Locating the ``missing'' baryons with extragalactic dispersion 
measure estimates, ApJ 780, L33 (2014) [arXiv:1309.4451] 

\bibitem{1712.01280} 
J.M. Shull, C.W. Danforth, The Dispersion of Fast Radio Bursts from a 
Structured Intergalactic Medium at Redshifts $z < 1.5$, ApJL 852, L11 (2018) 
[arXiv:1712.01280] 

\bibitem{1901.11051}
J.X. Prochaska and Y. Zheng, Probing Galactic haloes with fast radio bursts, 
MNRAS 485, 648 (2019) [arXiv:1901.11051] 

\bibitem{hu1} 
W. Hu, G. Holder, Model-Independent Reionization Observables in the CMB, 
Phys. Rev. D 68, 023001 (2003) [arXiv:astro-ph/0303400] 

\bibitem{hu2} 
M.J. Mortonson, W. Hu, Model-independent constraints on reionization from 
large-scale CMB polarization, ApJ 672, 737 (2008) [arXiv:0705.1132] 

\bibitem{shull12} 
J.M. Shull, B.D. Smith, C.W. Danforth, 
The Baryon Census in a Multiphase Intergalactic Medium: 30\% of the 
Baryons May Still Be Missing, ApJ 759, 23 (2012) [arXiv:1112.2706] 

\bibitem{snbias} 
E.V. Linder, A. Mitra, Photometric Supernovae Redshift Systematics 
Requirements, Phys. Rev. D 100, 043542 (2019) [arXiv:1907.00985] 

\bibitem{gwbias} 
R.E. Keeley, A. Shafieloo, B. L'Huillier, E.V. Linder, Debiasing Cosmic 
Gravitational Wave Sirens, MNRAS 491, 3983 (2020) 
%https://doi.org/10.1093/mnras/stz3304, 
[arXiv:1905.10216] 

\bibitem{ufa} 
S. Li, G.F. Smoot, B. Grossan, A.W.K. Lau, M. Bekbalanova, M. Shafiee, 
T. Stezelberger, Program objectives and specifications for the Ultra-Fast 
Astronomy observatory, Proc. SPIE 11341, 113411Y (2019) 
[arXiv:1908.10549] 

\end{thebibliography}
\end{document}